\documentclass{bmcart}
\usepackage{cite}
\usepackage[utf8]{inputenc} 

\usepackage{graphicx}

\startlocaldefs
\endlocaldefs

\begin{document}

\begin{frontmatter}

\begin{fmbox}
\dochead{Research}


\title{Efficient quantum simulation of fermionic and bosonic models in trapped ions}


\author[
   addressref={aff1},               
]{\inits{LL}\fnm{L.} \snm{Lamata}}
\author[
   addressref={aff1},
]{\inits{AM}\fnm{A.} \snm{Mezzacapo}}
\author[
   addressref={aff1},
]{\inits{JRS}\fnm{J.} \snm{Casanova}}
\author[
   addressref={aff1,aff2},
]{\inits{JRS}\fnm{E.} \snm{Solano}}

\address[id=aff1]{
  \orgname{Department of Physical Chemistry, University of the Basque Country UPV/EHU}, 
  \street{Apartado 644},                     %
  \postcode{E-48080}                                
  \city{Bilbao},                              
  \cny{Spain}                                    
}
\address[id=aff2]{%
  \orgname{IKERBASQUE, Basque Foundation for Science},
  \street{Alameda Urquijo 36},
  \postcode{48011}
  \city{Bilbao},
  \cny{Spain}
}



\end{fmbox}


\begin{abstractbox}

\begin{abstract} 

We analyze the efficiency of quantum simulations of fermionic and bosonic models in trapped ions. In particular, we study the optimal time of entangling gates and the required number of total elementary gates. Furthermore, we exemplify these estimations in the light of quantum simulations of quantum field theories, condensed-matter physics, and quantum chemistry. Finally, we show that trapped-ion technologies are a suitable platform for implementing quantum simulations involving interacting fermionic and bosonic modes, paving the way for overcoming classical computers in the near future.

\end{abstract}



03.67.Ac, 03.67.Lx, 37.10.Ty

\end{abstractbox}
%

\end{frontmatter}



\section{Background}
\label{Introduction}

Quantum simulation is one of the most promising fields in quantum information science. Feynman already pointed out in 1982~\cite{Feynman82} that a controllable quantum platform could simulate the dynamics or static properties of another quantum system exponentially faster than classical computers. Since then, this hypothesis has been demonstrated~\cite{Lloyd96}, and important theoretical and experimental work followed~\cite{Buluta09,Wunderlich09,SchneiderPorrasSchaetz}. Furthermore, quantum simulators establish analogies between previously unconnected fields, and have as a main aim to overcome classical computers.

Many proposals and experimental realizations of quantum simulations in a broad variety of platforms have been put forward, as for example spin systems~\cite{Jane,Porras,Friedenauer08,Kim10}, the Bose-Hubbard model implemented with cold atoms~\cite{Greiner02}, quantum chemistry~\cite{Lanyon10} and quantum statistics~\cite{Matthews,Sciarrino1,Sciarrino2} simulated with photonic systems, condensed matter models with Rydberg atoms~\cite{Weimar10}, relativistic quantum mechanics~\cite{Lamata07,Gerritsma1,Casanova1,Gerritsma2}, quantum field theories~\cite{CasanovaQFT}, and the lattice Schwinger model~\cite{Hauke}. On the other hand, quantum simulations of fermionic and bosonic systems in trapped ions have been recently proposed~\cite{CasanovaFermions}. Therefore, it is timely to study the experimental requirements needed, to assess the feasibility of the proposal and to compare it with other implementations.
  
 In this article, we analyze the necessary resources to implement a quantum simulation of fermions and bosons with trapped ions~\cite{CasanovaQFT,CasanovaFermions,MezzacapoHolstein,QChem}. We show that the methods developed for simulating fermionic and bosonic systems with ions can save a large amount of resources in terms of gates with respect to other platforms. This demonstrates that trapped ions are a promising quantum technology for a wide variety of quantum simulations, including high energy physics, condensed matter, or quantum chemistry.
 
\begin{figure}[h!]
\includegraphics[width=0.95\hsize]{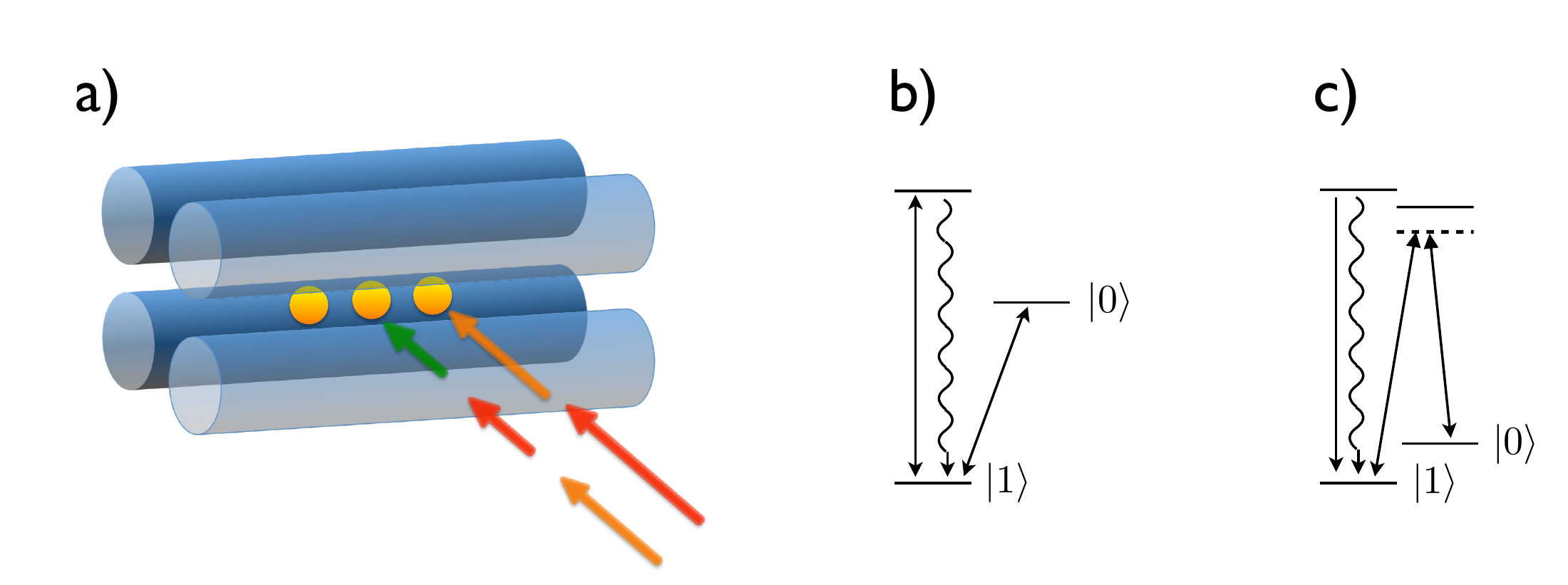}\caption{\csentence{Quantum simulations with trapped ions.} {a) Scheme of a  rf Paul trap for digital-analog quantum simulations.  Energy-level scheme for optical qubits (b), and rf qubits (c). In addition to the qubit levels $|0\rangle$ and $|1\rangle$,  a third level is used for qubit readout.}} \label{innsbrucktrap}
\end{figure}

Trapped-ion systems are one of the most advanced technologies for implementing quantum information protocols. Ions are charged particles that  can be confined either in Penning traps~\cite{GabrielseRMP} or radio-frequency  (rf) Paul traps~\cite{Leibfried03}. The former uses electrostatic and magnetic fields, whereas the latter requires time-dependent fields to confine the ions in an effective harmonic potential. Here, we will focus on rf Paul traps, see Fig~\ref{innsbrucktrap}$a$. Two different kinds of qubits are currently employed, optical qubits and radio-frecuency qubits (rf qubits). In the first ones, see Fig.~\ref{innsbrucktrap}$b$, two internal metastable electronic levels corresponding to an optical transition are used to encode the qubit. In the second ones, see Fig.~\ref{innsbrucktrap}$c$, a third level is used to mediate a two-photon transition between the hyperfine or Zeeman electronic levels of the qubit. 

Via sideband cooling, the ionic motional modes are able to reach their ground state, which is commonly used as a quantum bus to perform two-qubit gates between any pair of ions in a string. Finally, using resonance fluorescence by means of a cyclic transition, quantum nondemolition measurements of the qubit can be realized. Fidelities of state preparation, single- and two-qubit gates, and qubit measurement, are currently above 99\%~\cite{Leibfried03}.

The basic Hamiltonian describing the coupling of a two-level cold ion with a laser beam is ($\hbar=1$)
\begin{eqnarray}
\mathcal{H} = && \!\! \frac{\omega_0}{2}\sigma_z +  \nu a^{\dag}a + \Omega(\sigma_++\sigma_-) \\ \nonumber && \times \big( \exp [i(kz-\omega_lt+\phi)] + \exp [-i(kz-\omega_l t + \phi)] \big) ,
\end{eqnarray}
where $\sigma_\pm$ and $\sigma_z$ are  Pauli matrices associated with the ionic internal levels, $a$ ($a^\dag$) is the annihilation (creation) operator of the corresponding motional mode, $\omega_0$ is the frequency of the internal ionic transition, $\nu$ is the frequency of the trap, $\omega_l$ is the frequency of the laser field drive, $\phi$ is the laser phase, $k$ is the laser wave vector, and $\Omega$ is the Rabi frequency associated with the ion-laser coupling.

Transforming into an interaction picture with respect to the internal and motional free-energy term $\frac{\omega_0}{2}\sigma_z +  \nu a^{\dag}a$, and after application of the optical rotating-wave approximation, one obtains
\begin{eqnarray}
\mathcal{H}^{I} =  \Omega \sigma_+ e^{-i(\Delta t-\phi)} \exp[i \eta ( a e^{-i \nu t} + a^\dag e^{i \nu t})] + {\rm H.c.} , \label{BasicHIonsLaser}
\end{eqnarray}
where $\Delta$ is the laser detuning with respect to the internal ionic transition, $\eta=k z_0$ is  the so called Lamb-Dicke parameter, where $z_0 = \sqrt{1 / 2m \nu}$ is the ground state width of the motional harmonic oscillator mode.

In the Lamb-Dicke regime, namely, $\eta\sqrt{\langle (a+a^\dag)^2\rangle}\ll 1$, Eq.~(\ref{BasicHIonsLaser}) can be expressed as
\begin{eqnarray}
\mathcal{H}^I  =  \!\! \Omega[\sigma_+e^{-i(\Delta t-\phi)} + i \eta \sigma_+ e^{-i (\Delta t-\phi)} (a e^{-i \nu t}+a^\dag  e^{i \nu t}) + {\rm H.c.}] .
\end{eqnarray}

By choosing different internal vibrational transitions appropriately changing the laser detuning, $\Delta$, one can obtain the three basic interactions in trapped-ion technology. Namely, the carrier interaction ($\Delta=0$),
\begin{eqnarray}\label{carrier}
\mathcal{H}^I_c=\Omega(\sigma_+e^{i\phi}+\sigma_-e^{-i\phi}),
\end{eqnarray}
the red-sideband interaction ($\Delta = -\nu$),
\begin{eqnarray}\label{rsideband}
\mathcal{H}^I_r=i\eta\Omega(\sigma_+a e^{i\phi}-\sigma_- a^\dag e^{-i\phi}),
\end{eqnarray}
and the blue-sideband interaction ($\Delta = + \nu$),
\begin{eqnarray}\label{bsideband}
\mathcal{H}^I_b=i\eta\Omega(\sigma_+a^\dag e^{i\phi}-\sigma_- a e^{-i\phi}).
\end{eqnarray}

One can also take into account several laser drivings acting upon different ions in a string. In this situation, one can express the basic interactions as
\begin{eqnarray}\label{carriers}
\mathcal{H}^I_{c,j}&=&\Omega(\sigma^j_+e^{i\phi}+\sigma^j_-e^{-i\phi}),\nonumber\\
\mathcal{H}^I_{r,j,k}&=&i\eta\Omega(\sigma^j_+ a_ke^{i\phi}-\sigma^j_{-}a_k^{\dag}e^{-i\phi}),\nonumber\\
\mathcal{H}^I_{b,j,k}&=&i\eta\Omega(\sigma^j_{+} a_k^{\dag}e^{i\phi}-\sigma^j_{-}a_ke^{-i\phi}) ,
\end{eqnarray}
where $\sigma_{\pm}^j$ and  $a^{\dag}_k(a_k)$ are the raising and lowering operators of the $j$-th ion and the creation(annihilation)  bosonic operators of the $k$-th vibrational mode, respectively.

By appropriately combining the interactions appearing in  Eq.~(\ref{carriers}) one may obtain the basic single and two-qubit gates necessary for universal quantum computing. Prototypical cases of two-qubit gates that can be realized in trapped ions are: the Cirac-Zoller gate~\cite{CiracZoller95}, corresponding basically to a controlled-NOT (CNOT) gate, and the  M\o lmer-S\o rensen (MS) gate~\cite{MolmerSorensen99}, that is the basic building block for our quantum simulations of fermions and bosons in trapped ions.

 The structure of the article is as follows. In Section~\ref{FermionsBosons}, we summarize the method for simulating fermionic systems in trapped
 ions introduced in Ref.~\cite{CasanovaFermions}, and we propose a novel approach with an ultrafast gate that may speed up the implementation of the method. In Section~\ref{Feasibility} we assess the efficiency of the method in terms of the number of elementary gates and realization time, and show that it can be highly advantageous as compared to other platforms. Finally, we give our conclusions in Section~\ref{Conclusions}. 

\section{Results and discussion}
\subsection{Fermionic and bosonic models in trapped ions}
\label{FermionsBosons}
Interacting fermionic and bosonic systems are ubiquitous in physics. They appear as effective models in condensed-matter physics and quantum chemistry, constituting also   the natural language in which  quantum field theories are analyzed.
The numerical computation of interacting fermionic and bosonic models is, in general, a hard problem due to the fast growth of the Hilbert space dimension with the number of modes~\cite{Feynman82}.  The  use of numerical methods such as quantum Monte Carlo is not always possible due to the so-called sign problem. In this sense, quantum simulations appear as the technique that will allow us to calculate the time evolution of  interacting fermionic and bosonic  theories in an optimized way~\cite{Lloyd96}. 

\subsubsection{Efficient implementation of fermionic dynamics}

\begin{figure}[h]
\includegraphics[width=0.8\hsize]{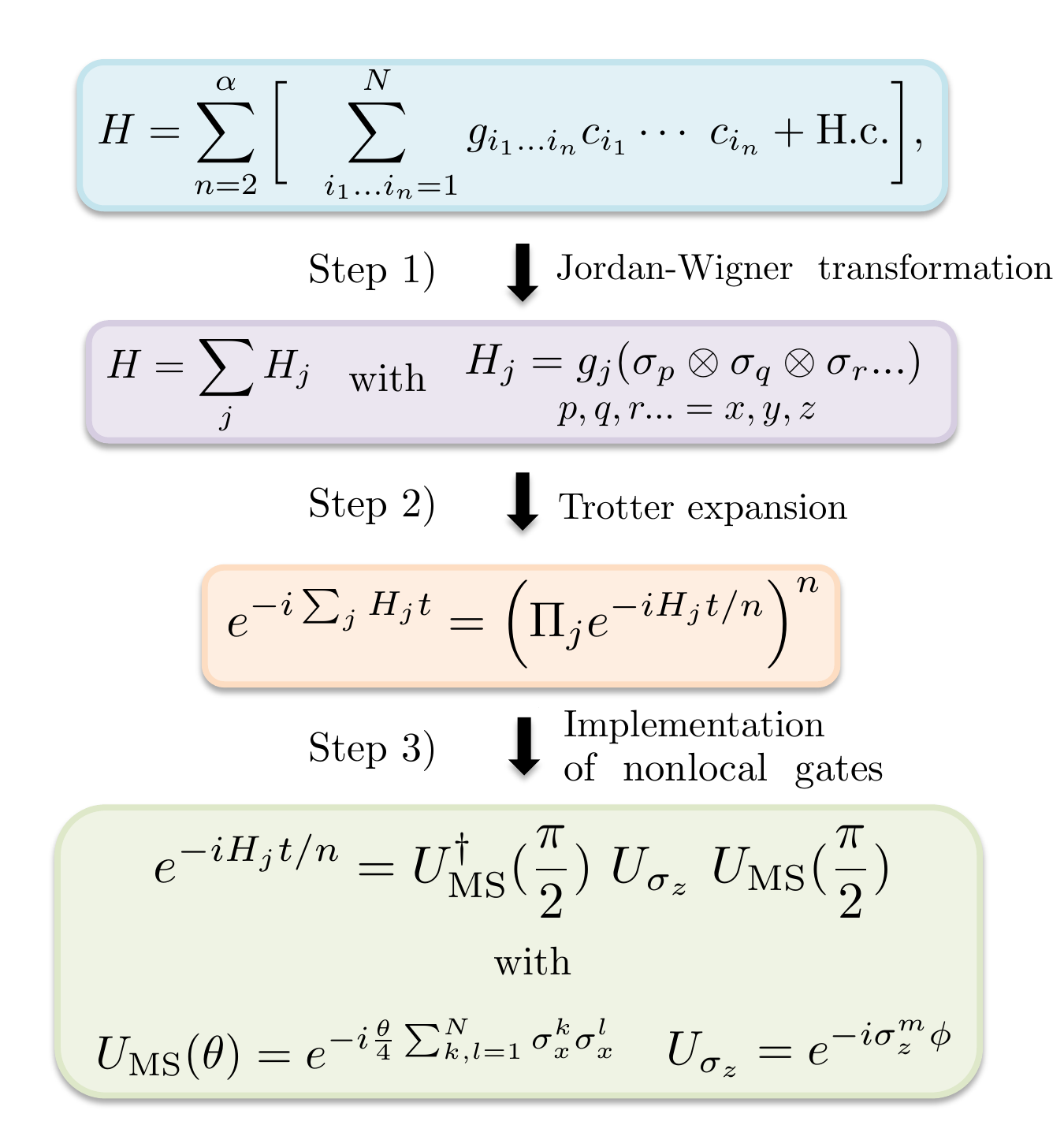}\caption{\csentence{Scheme of the proposed three-step protocol.} $1)$ Mapping of the Hamiltonian $H$, where each  $c_{i_k}$ corresponds to a  fermionic operator, $N$ is the total number of fermionic modes,  and $\alpha$ is the highest order of the many-body interaction, using the Jordan-Wigner transformation, to a sum of tensor products of  spin operators. $2)$ Trotter expansion of the resulting evolution. $3)$ Implementation of each nonlocal evolution, up to local rotations, with a sequence of two MS gates and a local gate. We point out that the use of local rotations allows one to obtain an arbitrary fermionic Hamiltonian. While the MS gates correspond to an interaction upon all the ions, the intermediate one is a local rotation applied on just one ion. To simulate a dynamics of fermions coupled to bosons, one would substitute the local gate $U_{\sigma_z}$ by $\exp[-i\phi\sigma_z^m(a+a^\dag)]$, resulting in a linear coupling of bosons to a quadratic fermionic operator.}\label{fig:protocol}
\end{figure}

Here, we show how a linear Paul trap can efficiently encode the dynamics of interacting fermionic and bosonic systems using digital-analog techniques~\cite{CasanovaFermions}. Our method consists of three steps, see Fig.~\ref{fig:protocol}. Firstly, we map the Hamiltonian $H$ of  $N$ interacting fermionic modes, via the Jordan-Wigner transformation, to a sum of nonlocal spin operators. The second step consists in decomposing the  evolution operator associated with $H$ via a Trotter expansion. This yields products of exponentials, each of them associated with one of the nonlocal spin operators appearing in the Hamiltonian. Finally, we implement  these exponentials, up to local rotations, on a set of $N$ two-level ions with a small number of laser pulses, by means of two M\o lmer-S\o rensen gates and a local gate. These three steps produce an efficient protocol  employing just polynomial resources. A trapped ion implementation  of the presented method will be  able to simulate nonlinear and long-range fermionic and bosonic interactions for two-dimensional and three-dimensional problems. The reason for this is the fact that the time evolution associated with the nonlocal spin operators, which contains a large number of Pauli operators, can nevertheless be efficiently realized (see the third step in Fig~\ref{fig:protocol}). For about~$\sim 40$ or more particles, one could already overcome classical computers in a fermionic quantum simulation. The requirements for efficient quantum simulation of fermionic or spin models may also be developed in superconducting circuits as recently shown in Refs.~\cite{MezzacapoMS,HerasSpins,MezzacapoRabi}.

\subsubsection{Efficient implementation of fermions interacting with bosons}

The same  protocol presented in Fig.~\ref{fig:protocol} can be extended in order to include bosonic modes in the formalism. In this sense, the only requirement corresponds to the replacement of  the gate $e^{-i \phi \sigma_z^m}$ by $e^{-i\phi \sigma_z^m  (a+a^{\dag})}$ in the third step of the algorithm.  The possibility to implement this kind of interactions in a quantum simulator generalizes the kind of theories to be simulated. Some examples are the Holstein model~\cite{Holstein59,MezzacapoHolstein}  in condensed matter physics or nontrivial extensions of quantum chemistry including the molecular vibronic degrees of freedom. Furthermore, the inclusion of bosonic modes  paves the way to the implementation of quantum field theories in trapped ions~\cite{CasanovaQFT}. Quantum field theories are among the deepest theories of nature describing the behavior of fundamental particles.  For example, quantum electrodynamics is the theory governing the interaction between fermionic charged particles  through the electromagnetic field. A simplified version of the Hamiltonian describing the interaction between fermions and bosons, when restricted to the case of $1+1$ dimensions and scalar particles, can be written as~\cite{Peskin}
\begin{equation}\label{qed}
H= g\int dx \psi^{\dag}(t, x)\psi(t, x) A(t, x).
\end{equation}
Here, the fermionic field $\psi$ is
\begin{equation}
\psi(t, x)=\frac{1}{\sqrt{2\pi}}\int dp(b_p e^{-i\omega_p t}e^{ipx}  + d^{\dag}_p e^{i\omega_p t}e^{-ipx})
\end{equation}
with $b_p$ $(d_p)$ fermionic (antifermionic) operators obeying the anticommutation rules $\{b_p, b^{\dag}_{p'} \}=\delta(p-p') $ and $\{d_p, d^{\dag}_{p'} \}=\delta(p-p') $. The bosonic electromagnetic field  $A(t,x)$ is
\begin{equation}
A(t, x) =\frac{1}{\sqrt{2\pi}}\int dk(a_k e^{-iw_k t} e^{ikx} + a^{\dag}_k e^{i\omega_k t} e^{-ikx}),
\end{equation}
where $a_k$ and $a^{\dag}_k$ obey $[a_k, a_{k'}^{\dag}] = \delta(k-k')$.

Here, several approximations can be considered. For instance, an adequate discretization in the number of fermionic and bosonic modes  will make the  implementation feasible with current technology. We point out that the kind of terms that will arise from Hamiltonian~(\ref{qed}) simply correspond to  interactions between fermions and fermions with bosons, and each of them is implementable with the previously commented techniques. In the spin language, the different terms appearing will be nonlocal spin operators coupled linearly to the position operator of the bosons. The interaction between fermions will arise through the coupling to the bosons, that will act as mediators.

\subsubsection{Optimization of entangling gates for implementing fermions and bosons in trapped ions}

One possibility to improve the gate time in our protocol for simulating fermionic and bosonic systems in trapped ions is the following. It can be realized that the Hamiltonian of the MS gate used in our protocol, 
\begin{equation}
H_{\rm MS}= g\sum_{j,k=1}^N \sigma_x^j\sigma_x^k, 
\end{equation}
may be substituted by the Hamiltonian 
\begin{equation}
H_{\rm UMQ}= g\sigma_x^1\sum_{j=2}^N \sigma_x^j,
\end{equation}
associated with a dynamics that we name {\it Ultrafast Multi-Qubit} (UMQ) gate for reasons that will be clarified later. Label 1 denotes here the ion acted upon by the local gate applied between the two MS gates. Doing this substitution the protocol remains exactly the same, given that  the terms not involving ion 1 cancel out after the two MS gates. Notice that the label 1 can refer to any spin which appears in the nonlocal spin interaction. Accordingly, one may just consider the UMQ gate for our protocol instead of the MS gate. That is, one would implement the UMQ gate pair by pair. This can be a significant advantage, given that unitary evolutions associated with Hamiltonians of the kind $\sigma_x^1\sigma_x^j$ can be done in trapped ions with resonant gates~\cite{GarciaRipollGate,DuanGate}, saving orders of magnitude in the gate time with respect to dispersive gates like the MS gate.  Current proposals consider applying the resonant gates to neighboring ions, and to couple distant ions one can employ shuttling~\cite{GarciaRipollGate,DuanGate}. We consider a setup analogous to the one in Ref.~\cite{Kielpinski}. The shuttling time will depend on the specific experiment and the kind of protocol, either adiabatic or making use of reverse engineering techniques. Thus, for say 10 ions, one will need 10 resonant gates in the new protocol per MS gate in the old protocol, to implement the UMQ gate. However, given that the resonant gates can be done about 3 orders of magnitude faster, one may reduce the total protocol time by 2 orders of magnitude. For the relevant case of 40 ions, which would simulate 40 fermionic modes, one may reduce the simulator time a factor of 25 using the UMQ gate with respect to the MS-based protocol. Instead of shuttling the ions, a further possibility is to couple distant sites via swap gates with just a linear resource overhead~\cite{DuanGate}.

\subsection{Analysis of quantum-simulated dynamics}
\label{Feasibility}

We assess now the feasibility of our protocols for implementing fermionic and bosonic dynamics in trapped ions by giving illustrative examples. We calculate the number of entangling gates needed when using MS gates, and compare it with other implementations that use Controlled-NOT (CNOT) gates~\cite{Nielsen00}. We show that in trapped ions one may reduce the number of necessary multiqubit gates by almost one order of magnitude. 

\subsubsection{Fermion models in trapped ions}
\label{sec:2}

\begin{figure}[t!]
\includegraphics[width=0.8\hsize]{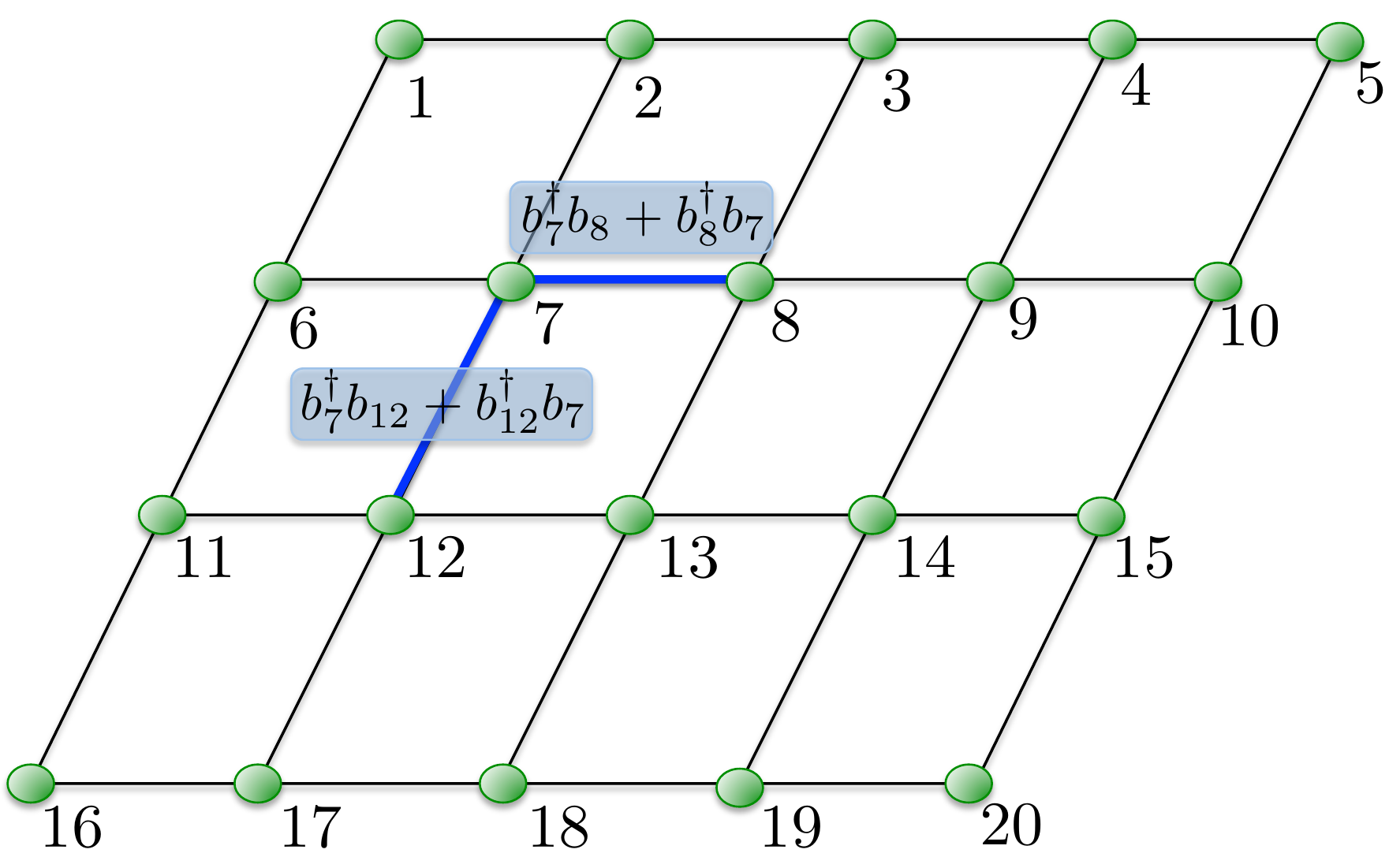}\caption{\csentence{Scheme of the lattice for a quantum simulation of the Hubbard model with 40 fermionic modes.}}\label{fig:plaquette}
\end{figure}

We consider a trapped-ion quantum simulation of a $5 \times 4$ site fermionic lattice with two modes per site ($\uparrow, \downarrow$), see Fig.~\ref{fig:plaquette}.  We propose to map this lattice onto a 40-ion string in a Paul trap. Developments of an ion trap allowing for up to 1000 ions in a string, for quantum simulations, and quantum interfaces, have been recently published~\cite{VuleticChuangSurfaceTrap,LamataInterface}.  The total number of fermionic modes in our proposed system, i.e., 40, makes it be in the limit of the computational power of classical computers. Thus, this quantum simulator is one example of a relevant physical problem that trapped ions could solve efficiently while being hard for classical devices. 

We compare now the total number of entangling gates needed both in our trapped ion implementation, making use of MS gates, and in other quantum optics platforms, that may employ CNOT gates. We point out that to implement an $N$-body spin operator, one typically needs $2N$ CNOT gates, while, with our protocol, in general one just needs 2 MS gates. We consider the Hubbard model in the lattice shown in Fig.~\ref{fig:plaquette}. The system Hamiltonian is 
\begin{equation}
H= w\sum_{\langle i, j \rangle,\sigma} b^\dag_{i,\sigma}b_{j,\sigma}+{\rm H.c.}+U\sum_{i} b^\dag_{i,\uparrow}b_{i,\uparrow}b^\dag_{i,\downarrow}b_{i,\downarrow},\label{Hubbard40modesmodel}
\end{equation}
where $\sigma=\uparrow$ or $\downarrow$. Let us now make a counting of the number of gates needed for this quantum simulation. The first kind of terms in Eq.~(\ref{Hubbard40modesmodel}) that we consider are tunneling terms in a row, see Fig.~\ref{fig:plaquette}. After applying the Jordan-Wigner transformation, $b^\dag_{1\downarrow}=I^{\otimes 39}\otimes \sigma_+^1$, $b^\dag_{1\uparrow}=I^{\otimes 38}\otimes\sigma_+^2\otimes \sigma_z^1$,..., we have, e.g., $b^\dag_{7,\uparrow}b_{8,\uparrow}+{\rm H.c.}=(\sigma_x^{14}\otimes\sigma_z^{15}\otimes\sigma_x^{16}+\sigma_y^{14}\otimes\sigma_z^{15}\otimes\sigma_y^{16})/2$. This kind of term employs 4 MS gates (two per summand), and given that we have 16 of these terms, with two spins per site, the total number of entangling gates needed is 128 MS gates. The second kind of terms in Eq.~(\ref{Hubbard40modesmodel})  are the tunneling terms in a column which are nonlocal in the spin-operator encoding. For example, the term  $b^\dag_{7,\downarrow}b_{12,\downarrow}+{\rm H.c.}=-(1/2)(\sigma_x^{13}\otimes\sigma_z^{14}\otimes...\otimes\sigma_z^{22}\otimes\sigma_x^{23}+\sigma_y^{13}\otimes\sigma_z^{14}\otimes...\otimes\sigma_z^{22}\otimes\sigma_y^{23})$. The number of MS gates of this kind of term is $15\times 2\times 2\times 2=120$. Finally, the last type of term in Eq.~(\ref{Hubbard40modesmodel}) is the onsite Coulomb repulsion, $U\sum_{i} b^\dag_{i,\uparrow}b_{i,\uparrow}b^\dag_{i,\downarrow}b_{i,\downarrow}$.  In the spin language we have, e.g., $b^\dag_{1,\uparrow}b_{1,\uparrow}b^\dag_{1,\downarrow}b_{1,\downarrow}=|0\rangle_2\langle 0|\otimes|0\rangle_1\langle 0|$, where $|0\rangle_i$ is the excited state of ion $i$ (see Fig.~\ref{innsbrucktrap}), which requires just one entangling gate,  with Hamiltonian $H\propto\sigma_z^1\otimes\sigma_z^2$. The number of two-qubit gates associated with this kind of term is 20. Summarizing, the number of entangling gates, including MS gates, for implementing the Hubbard model in our 40-mode lattice is 268 gates per Trotter step. For 10 Trotter steps, this will give a total number of 2680 gates.  

We want to compare now the previous calculation with fermionic implementations using digital simulations in other quantum platforms. In those cases in which the MS gate is not available, and one has to employ, e.g., CNOT gates, for each nonlocal spin operator composed of $N$ Pauli matrices one needs $2N$ CNOT gates~\cite{Nielsen00}, instead of 2 MS gates. Accordingly, the number of CNOT gates for the tunneling terms in a row is $16\times 2\times 6=192$ for each spin, 384 in total. Moreover, the number of CNOT gates for the tunneling terms in a column is $15\times 2\times 22=660$ per spin, 1320 in total. The number of entangling gates for the onsite Coulomb repulsion is still the same as before, 20. Accordingly, the total number of entangling gates per Trotter step is in this case 1724. For 10 Trotter steps, the total number of gates in implementations making use of CNOT gates and not of MS gates is of about 17000. This is almost one order of magnitude larger than for the trapped ion case. Accordingly, one realizes that trapped-ion implementations, making use of MS gates, are a highly versatile tool for implementing a fermionic dynamics. Even though one can simulate free fermions efficiently with cold atoms, given that in this case the atoms can be fermionic, it is much more difficult to have nonlinearities and interactions among them than with our trapped-ion methods. Accordingly, we consider our protocol for implementing fermions in trapped ions as complementary to cold atoms and optical lattices.

We make now a rough estimate of the total time of the protocol: for a MS gate time of about 20 $\mu$s, and for $\sim$2500 entangling gates, assuming the single qubit gates will contribute a small fraction of the total evolution period, this will give a total protocol time of about 50 ms. This is on the order of magnitude of the decoherence time in some current experiments for digital quantum simulators~\cite{Lanyon11}. Accordingly, near-future improvements may allow to reach similar numbers of gates in the protocol inside the coherence time. For a large number, $N$, of ions in the trap, the protocol may take a longer time for nonresonant gates due to the reduced ion-phonon coupling, that scales as $1/\sqrt{N}$. The time of the MS gate will then increase by a factor of $N/2$, for $N$ ions with respect to 2, i.e., for 40 ions a factor of 20. This will give a total time of the protocol of 1 s, which is, on the other hand, about the decoherence time for some quantum-simulation experiments with trapped ions, e.g., with Ytterbium~\cite{Kim}. A further possibility to reduce the time of the protocol, and in consequence increase the number of gates that may be realized in an experiment, is to substitute the standard dispersive MS gates by a UMQ gate based on resonant entangling gates, like in Refs.~\cite{GarciaRipollGate,DuanGate}, as we pointed out in Sec.~\ref{FermionsBosons}. Given that we just need to apply $N$ of these gates per MS gate as shown in previous section, and we can reduce the time of each gate by about 3 orders of magnitude, for $N=40$ one could reduce the time of the protocol a factor 25, well below the decoherence time. An optimization of resonant entangling gates for simulating fermions in trapped ions may significantly improve these resources.  A further issue is the fact that each single or two-qubit gate will have a finite error that will accumulate when applying many of them. On the other hand, recent estimations point to the fact that MS gates may be done in the near future with fidelity errors smaller than 10$^{-4}$~\cite{KirchmairMSError}. If this is the case, reaching about thousands of gates in a single quantum simulation experiment with trapped ions, without error correction, may be a realistic possibility.

\subsubsection{Fermions coupled to bosons in trapped ions}
\label{sec:2}
Including bosons in the fermionic trapped-ion simulation, as explained in Sec.~\ref{FermionsBosons}, may give a significant computational power increase. From a numerical point of view, when one considers phonon generation, the Hilbert space dimension of the simulated system can grow fast even with a small number of fermionic sites. Notice that the dynamics of a bosonic Hilbert space can be approximated with truncation at some level of the number of bosonic quanta. Therefore, the amount of computational resources that one needs in order to obtain a given fidelity for the quantum state of the simulation increases with the amount of bosonic excitations. Given that the bosonic part is analog in our formalism, just with 10 ions, 10 motional modes, and up to 7 phonons per mode, one would reach a total Hilbert space dimension of $2^{40}$, which is at the limit of the fastest classical computers. An example of fermionic-bosonic simulation is the Holstein model~\cite{MezzacapoHolstein}. The Holstein Hamiltonian is given~by
\begin{equation}
H=h\sum_{i=1}^{N-1}(b_{i}^{\dagger}b_{i+1}+{\rm H.c.})+g\sum_{i=1}^{N}(a_{i}+a_{i}^{\dagger})b^{\dagger}_{i}b_i+
\omega_{0}\sum_{i=1}^{N}a_{i}^{\dagger}a_{i}\label{HolsteinHam}.
\end{equation}
 Here, $b_{i} (b_{i}^{\dagger})$ is the annihilation (creation) operator at the fermion site $i$, and $a_{i}(a_{i}^{\dagger})$ is the phonon annihilation (creation) operator at site $i$. The parameters $h$, $g$ and $\omega_0$ are respectively a nearest-neighbor (NN) site hopping for the fermions, fermion-phonon coupling and phononic free energy. The model describes a fermion-phonon correlated system, which has been proven to be of great importance for a large number of solid state systems. In condensed-matter systems, the correlation between the presence of fermions in a lattice and distortions of the latter can produce the creation of polarons: fermions and phonons can no more be regarded as independent particles. The lattice distortion surrounding the fermion can have different size depending on the intensity of the fermion-phonon interaction. For strong coupling, the fermions can become trapped, with interesting changes of global transport properties. 
 
 We can map via Jordan-Wigner transformation the original Hamiltonian into a coupled spin-boson system. The mapped Hamiltonian has the form
\begin{eqnarray}
\nonumber H=h\sum_{i=1}^{N-1}(\sigma^{i}_{+}\sigma^{i+1}_{-}+{\rm H.c.})+g\sum_{i=1}^{N}(a_{i}+a_{i}^{\dagger})\frac{(\sigma^{i}_{z}+1)}{2}
+\omega_{0}\sum_{i=1}^{N}a_{i}^{\dagger}a_{i}.
\end{eqnarray}
To implement an $N$-site Holstein lattice in a trapped-ion chain, we make use of $N+1$ ions. The first $N$ ions will encode the dynamics of the fermionic sites, while the last passive ion is used to displace the motional modes. The first term can be rewritten as $\frac{h}{2}\sum_{i=1}^{N}(\sigma^{i}_{x}\sigma^{i+1}_{x}+\sigma^{i}_{y}\sigma^{i+1}_{y})$. One can implement in a single MS gate the interactions  $\frac{h}{2}\sigma^{i}_{x}\sigma^{i+1}_{x}$ and $\frac{h}{2}\sigma^{i}_{y}\sigma^{i+1}_{y}$, for each couple of nearest-neighbor ions. One additional, combined red-blue sideband interaction can implement the interaction of each ion of the chain with a particular normal mode $a_i$. Finally, driving the passive ion in the chain, one can realize the free displacement of the motional modes in the model.

 In summary, with ten sites, in order to perform one Trotter step, one will need $9\times2=18$ MS local gates to implement the NN fermionic interactions, $10$ red-blue detuned sidebands for having the fermionic-bosonic interactions, and $10$ pulses to couple each normal mode to the passive ion. Therefore, a number of 38 gates per Trotter step, and considering 10 Trotter steps, gives a total number of 380 gates, which is foreseeable in the near future. 
 
In general, for the total $N$-mode Hamiltonian, one will need $2(N-1)$ MS gates and $2N$ red-blue detuned sidebands per Trotter step. This is a number of gates linear in $N$. This should be compared with the resources needed for a classical simulation of this dynamics: when the number of bosonic excitations per mode exceeds 7, the associated Hilbert space will have a dimension larger than $2^{4N}$, overcoming the capacities of classical computers for even small $N$. Regarding the total time of the protocol, for 10 fermionic and bosonic modes, one will need 11 ions, and the MS gate will employ a time about 5 times longer than for just 2 ions, while the red and blue sidebands will take a time about 2 times longer. Accordingly, the time of the protocol for a typical dynamics will be of about 2 ms per Trotter step.

\subsubsection{Quantum chemistry problems in trapped ions}

We describe now how to deal with quantum chemistry problems within the approach presented in the previous sections~\cite{QChem}. Typical quantum chemistry problems involve the many-body interactions describing electrons and nuclei. A generic quantum chemistry Hamiltonian, $H = {T_e} + {V_e} + {T_N} + {V_N} + {V_{eN}}$, contains the kinetic energies of the electrons ${T_e} \equiv  - {\textstyle{{{\hbar ^2}} \over {2m}}}\sum\nolimits_i {\nabla _{e,i}^2}$ and nuclei ${T_N} \equiv  - \sum\nolimits_i {{\textstyle{{{\hbar ^2}} \over {2{M_i}}}}\nabla _{N,i}^2}$, and the electron-electron interactions ${V_e} \equiv \sum\nolimits_{j > i} {{e^2}} /\left| {{{\bf r}_i} - {{\bf r}_j}} \right|$, the nuclei-nuclei potential energy ${V_N} \equiv \sum\nolimits_{j > i} {{Z_i}{Z_j}{e^2}} /\left| {{{\bf R}_i} - {{\bf R}_j}} \right|$, and electron-nuclei interaction ${V_{eN}} \equiv  - \sum\nolimits_{i,j} {{Z_j}{e^2}} /\left| {{{\bf r}_i} - {{\bf R}_j}} \right|$, where we have used ${\bf r}$ and ${\bf R}$ to address electron and nuclei coordinates.
Our methods can be complemented with a classical variational optimization method. In this way, one can perform a kind of  quantum-assisted optimization, in order to approximate ground-state energies and the ground-state eigenvectors for electronic molecular problems. Additional bosonic degrees of freedom can be encoded in the motional normal modes of the trapped ions to simulate vibrational and rotational degrees of freedom of a quantum chemistry problem.
A generic quantum chemistry Hamiltonian, including two-body interactions in second-quantization formalism, reads
\begin{equation}\label{h_pq_h_pqrs}
{H_e}({\bf R}) = \sum\limits_{pq} {{h_{pq}}b_p^\dagger {b_q}}  + \frac{1}{2}\sum\limits_{pqrs} {{h_{pqrs}}b_p^\dagger b_q^ \dagger {b_r}{b_s}}.
\end{equation}
Here, $b_p$ ($b_p^{\dagger}$) are annihilation (creation) operators for the electronic molecular orbitals, $h_{pq}$ are the one-particle amplitudes obtained from the computation of single electron integrals of the kinetic energy of the electron and the nuclei-electron interaction. The coefficients $h_{pqrs}$ come from the two-particle electron-electron Coulomb repulsion. Namely, if ${\phi _p}\left( {\bf r} \right) \equiv \left\langle {\bf r} \right|  {p} \rangle$ is the coordinate wavefunction of mode $p$, one has explicitly that ${h_{pq}} \equiv  - \int {d{\bf r}} \phi _p^*\left( {\bf r} \right)\left( {{T_e} + {V_{eN}}} \right){\phi _q}\left( {\bf r} \right)$, and the two-body term ${h_{pqrs}} \equiv \int {d{{\bf r}_{1,2}}} \phi _p^*\left( {{{\bf r}_1}} \right)\phi _q^*\left( {{{\bf r}_2}} \right){{V_e}\left( {\left| {{{\bf r}_1} - {{\bf r}_2}} \right|} \right)}{\phi _r}\left( {{{\bf r}_2}} \right){\phi _s}\left( {{{\bf r}_1}} \right)$. To obtain this Hamiltonian one performs the Born-Oppenheimer approximation and chooses a finite basis of molecular orbitals to expand the single-body and two-body terms. To include the bosonic degrees of freedom in the description, one can go beyond the Born-Oppenheimer approximation and expand the term $V_{eN}$ linearly in the nuclei positions. Later, one substitutes these positions by the nuclei normal modes, computed in the Born-Oppenheimer zeroth order approximation, and obtains a linear bosonic term coupled to quadratic fermionic terms, that can account for nonadiabatic corrections to the Born-Oppenheimer solution that one can solve with our methods. To implement the dynamics of the Hamiltonian in Eq.~(\ref{h_pq_h_pqrs}) one can map it into a spin representation via Jordan-Wigner transformation, as described in Fig.~\ref{fig:protocol},
\begin{equation}\label{He_JWT}
{H_e}  \mathrel{\mathop{\kern0pt\longrightarrow} \limits_{{{\rm JWT}}}}  \sum\limits_{i,j,k...\in \left\{ {x,y,z} \right\}} {{g_{ijk...}}\left( {\sigma^1_i \otimes \sigma^2_j \otimes \sigma^3_k...} \right)}.
\end{equation} 
The associated dynamics can be simulated digitally with a Trotter expansion, as in Fig.~\ref{fig:protocol}, implementing single nonlocal terms as a combination of global M\o lmer-S\o rensen gates and local ion rotations. 

The mean value of the energy $\langle{H_e}\rangle$ can be retrieved by obtaining mean values over individual terms of the Hamiltonian. Specifically, the expectation value of products of Pauli matrices can be mapped onto single qubit expectation values through application of nonlocal qubit gates~\cite{QChem}. Finally, the eigenvalues of $H_e$ can be obtained using the phase estimation algorithm, starting from trial states obtained with the previous method via unitary coupled-cluster techniques~\cite{QChem}. By letting the system evolve for different times, with a digital decomposition, one can perform the phase estimation algorithm with efficient resources.

One can also think of including vibrational and rotational degrees of freedom. The nonlocal gate can be performed in order to entangle the internal state of the ions with the normal motional modes, e.g. with operations like $\left| {{\psi _\theta }} \right\rangle  \equiv {e^{ - i\theta (\sigma^1_i \otimes \sigma^2_j \otimes  \cdots ) \left( {a + {a^\dag }} \right)}}\left| \psi  \right\rangle$. On the one hand, one is able to compute the time evolution with fermionic-bosonic Hamiltonians, and, on the other hand, to map correlation functions of coupled bosonic-fermionic degrees of freedom into the state of one single ion~\cite{QChem}.
As described in Sec.~\ref{sec:2}, a trapped-ion quantum computer with tens of fermionic and bosonic modes, implemented with a few hundreds of gates, would be able to perform quantum simulations within coherence times. As shown in the previous section, one would easily overcome a classical computer, dealing with Hilbert space dimension~$\sim2^{40}$. Although a quantum chemistry simulation with chemical accuracy of the ground state energy of a large molecule would require many gates~\cite{Wecker}, including bosonic degrees of freedom can significantly reduce this number. Moreover, for computing time evolution of quantum chemistry dynamics of even small molecules with correlations, a quantum simulator will be required, due to the fast entanglement growth that prevents the use of classical methods. In Sec. \ref{sec:2} we have included specific examples of numbers of gates and total time of the protocols for specific fermion-boson Hamiltonians that can be extrapolated to the case of quantum chemistry dynamics.

Summarizing, we propose a hybrid quantum-classical simulation approach involving digital-analog methods for quantum chemistry.

\section{Conclusions}
\label{Conclusions}
We have analyzed the feasibility of efficient quantum simulations of fermionic and bosonic interacting models in trapped ions. We have shown that these quantum simulations are advantageous when compared with other quantum platforms and with classical computers. Accordingly, the next generation of experiments with trapped ions may produce useful insight in the fundamental properties of condensed-matter, high-energy physics, and quantum chemistry models.

\begin{backmatter}

\section*{Competing interests}
  The authors declare that they have no competing interests.

\section*{Author's contributions}
    All the authors contributed to the realization and writing of the manuscript. 

\section*{Acknowledgments}
\label{Acknowledgments}
We thank Rene Gerritsma, Ben Lanyon, Juan Carlos Retamal, and Guillermo Romero for useful discussions. We acknowledge support from Basque Government IT472-10, Spanish MINECO FIS2012-36673-C03-02, Ram\'on y Cajal Grant RYC-2012-11391, UPV/EHU UFI 11/55, as well as from PROMISCE, CCQED, and SCALEQIT European projects.

\end{backmatter}

\end{document}